\documentclass[5p,twocolumn]{elsarticle}

\usepackage{graphicx}
\usepackage{dcolumn}
\usepackage{pifont}
\usepackage{bm}
\usepackage{multirow}
\usepackage{amsmath}
\usepackage{float}
\usepackage{txfonts}
\usepackage[version=3]{mhchem}

\usepackage[usenames,dvipsnames]{xcolor}
\definecolor{blue}{RGB}{0,0,225}
\definecolor{cream}{RGB}{222,217,201}
\definecolor{red}{RGB}{225,0,0}

\newcommand{\scunit}{mA h g$^{-1}$}

\journal{Elsevier}

\begin{document}
\title{First-Principles Study on \ce{Na_{$x$}TiO2} with Trigonal Bipyramid Structures: An Insight into Sodium-Ion Battery Anode Application}

\author[kimuniv-m]{Song-Hyok Choe}
\author[kimuniv-m]{Chol-Jun Yu\corref{cor}}
\cortext[cor]{Corresponding author}
\ead{cj.yu@ryongnamsan.edu.kp}
\author[kimuniv-m]{Kum-Chol Ri}
\author[kimuniv-m]{Jin-Song Kim}
\author[kimuniv-m]{Un-Gi Jong}
\author[kimuniv-m]{Yun-Hyok Kye}
\author[kimuniv-m]{Song-Nam Hong}

\address[kimuniv-m]{Chair of Computational Materials Design, Faculty of Materials Science, Kim Il Sung University, Ryongnam-Dong, Taesong District, Pyongyang, Democratic People's Republic of Korea}

\begin{abstract}
Developing efficient anode materials with low electrode voltage, high specific capacity and superior rate capability is urgently required on the road to commercially viable sodium-ion batteries (SIBs).
Aiming at finding a new SIB anode material, we investigate the electrochemical properties of \ce{Na_{$x$}TiO2} compounds with unprecedented penta-oxygen-coordinated trigonal bipyramid (TB) structures by using the first-principles calculations.
Identifying the four different TB phases, we perform the optimization of their crystal structures and calculate their energetics such as sodium binding energy, formation energy, electrode potential and activation energy for Na ion migration.
The computations reveal that TB-I phase can be the best choice among the four TB phases for the SIB anode material due to relatively low volume change under 4\%~upon Na insertion, low electrode voltage under 1.0 V with a possibility of realizing the highest specific capacity of $\sim$335 \scunit~from fully sodiation at $x=1$, and reasonably low activation barriers under 0.35 eV at the Na content from $x=0.125$ to $x=0.5$. 
Through the analysis of electronic density of states and charge density difference upon sodiation, we find that the \ce{Na_{$x$}TiO2} compounds in TB phases change from electron insulating to electron conducting material due to the electron transfer from Na atom to Ti ion, offering the \ce{Ti^{4+}}/\ce{Ti^{3+}} redox couple for SIB operation.
\end{abstract}

\begin{keyword}
Titania \sep Sodium-ion battery \sep Anode \sep Trigonal bipyramid \sep First-principles
\end{keyword}

\maketitle

\section{Introduction}
Along with the depletion of fossil fuels and serious concerns on global warming, ever growing energy demand in human life and industry highlights an importance of exploiting sustainable and natural clean energy sources such as solar and wind power~\cite{Paris,Chu17nm}.
Due to the intermittent character of natural energy sources, the energy harvesting or conversion devices should be paired with large-scale energy storage systems including rechargeable batteries typically~\cite{Dunn11s}.
For the batteries to be used in the grid, their component materials such as electrode and electrolyte are necessary to be based on earth-abundant, environmentally benign and cost-effective resources.
These requirements make the relatively expensive lithium-ion batteries (LIBs) less appealing due to the high cost and uneven distribution of lithium resources, despite their successful applications mostly in portable electronics and often in electric vehicles during the past decades~\cite{Tarascon10nc,Goodenough10cm}.
Therefore, the scientific interest has been shifted into developing non-lithium, so-called ``post-lithium'' batterries with lower cost and similar or even better performance~\cite{Sawicki}.
In this context, sodium-ion batterries (SIBs) have attracted a great deal of attention as the most promising alternative to LIBs, due to their meeting the aforementioned requirements and the same ``rocking-chair'' operational mechanism~\cite{JYHwang17csr,Kundu,Yabuuchi14}.

Sodium, however, has a larger ionic radius than Li, which makes it difficult to find suitable electrode materials that can be de/intercalated with Na during charge/discharge process, rising the problems of cyclic stability and rate capability of battery.
For commercially viable SIBs, therefore, it is vital to develop stable and high-performance electrode materials~\cite{Shi17jmca,Wang15}.
Although many candidates have been identified as promising cathode materials of SIBs, only a limited number of materials has been reported on the anode side so far.
As the most widely used and standard anode material with a theoretical capacity of 372 \scunit~for LIBs, graphite was found to have a very low specific capacity of $\sim$35 \scunit~ in SIBs, because it is impossible for Na to form low-stage graphite intercalation compounds due to a big mismatch between the ionic radius of \ce{Na+} cation and the interlayer distance of graphite~\cite{Dresselhaus}.
Thus, Na co-intercalation with organic molecules into graphite has been devised, but the capacity was found to be still limited to $\sim$100 \scunit~\cite{Jache,yucj06,yucj14,yucj20}.
So non-graphitic carbonaceous materials have been investigated.
Hard carbon showed the significantly higher capacity of $\sim$300 \scunit~but with the shortcoming of low Coulombic efficiency~\cite{Balogun,Irisarri,Tsai}, while multilayered graphene and its composites have been proved to be able to be used as anode but suffering from the weak binding of Na to the graphene sheet, resulting in Na clustering and dendrite growth~\cite{Lin16nn,Bonaccorso15s,YXWang13c,Pan14ic,Zhou14ic}.
The graphene-like two-dimensional (2D) materials like transition metal dichalcogenides~\cite{Chhowalla13nm,Kang17jmca,Mortazavi14jps,Pumera14jmca} and oxides~\cite{Jiang} have been reported to have exceptionally high initial capacities based on conversion and alloying reaction of Na storage mechanism, such as 847 \scunit~for Na$_{3.75}$Sn~\cite{zhu}, 660 \scunit~for Na$_{3}$Sb~\cite{darwiche} and 386 \scunit~for \ce{Fe2O3}.
The amorphous red phosphor even shows the reversible capacity of over 2000 \scunit~\cite{qian}.
But these materials suffer from fast fading capacity due to extremely large volume expansion and agglomeration of reaction products, and also from safety issue due to the formation of metal plates during the conversion reaction.
In order to alleviate these problems, the state-of-the-art nano technologies have been utilized, but such high-tech employment again impedes the scale-up to mass production of these materials.

The intercalation type materials can easily suppress such problems over cyclic stability and safety issues due to their ``rocking-chair'' mechanism and thus further researches to find better Na-intercalation compounds based on non-carbon materials are still in progress.
\textit{Inter alia} \ce{TiO2} polymorph can intercalate \ce{Na+} cation reversibly and thus can be used as potential anode materials for SIBs~\cite{Stefano,Usui,Huang,Wu15,Perez,Alexandros,Shen14}.
The \ce{TiO2} nanoparticles with anatase~\cite{Stefano}, rutile~\cite{Usui} and monoclinic phase~\cite{Huang} were found to show the reversible capacities of $\sim$100 \scunit, but whether they are based on the rocking-chair mechanism is still controversial due to their narrow interstitial space.
Other \ce{TiO2} polymorph with more open structures have shown the higher reversible capacities of $\sim$150 \scunit, corresponding to electrochemical reversibility between \ce{Na_{0.5}TiO2} and \ce{NaTiO2} in O3-layered structure~\cite{Wu15}, and $\sim$280 \scunit~in microporous hollandite-type structure~\cite{Perez}.
For the latter case, however, only 85 \scunit~remained to be reversible, implying that its microporous structure still cannot afford to host more than 0.25 Na per one formula unit (f.u.).
Very recently, Ma {\it et al.}~\cite{Sicong} have identified a class of new microporous \ce{TiO2} crystal structures composed of unprecedented penta-oxygen-coordinated (\ce{TiO5}) trigonal bipyramid (TB) building blocks, with a large amount of data from global potential energy surface calculations of \ce{TiO2}.
Interestingly, these TB structures have one-dimensional (1D) microporous channels with large pore sizes of 5.6$-$6.7 \AA, which might be favorable for ion intercalation and migration, and so the authors performed the density functional theory (DFT) calculations, demonstrating that these compounds can be good anode materials for LIBs with the whole theoretical capcity of $\sim$335 \scunit~offered by \ce{Ti^{3+/4+}} redox couple.
These microporous crystal structures of \ce{TiO2} can also be utilized in SIBs and even in the case of emerging concept of multivalent-ion batteries(MVIB) using divalent (\ce{Ca^2+}, \ce{Mg^2+}, \ce{Zn^2+}) or trivalent (\ce{Al^3+}) ions as charge carrier, although these materials are not yet reported to be synthesized in experiment.

In this work, we propose that \ce{TiO2} crystals with TB structures are highly promising anode material for SIBs based on the first-principles DFT calculations.
Our calculation reveals that this unique microporous crystal structure can afford the exceptionally high reversiblie capacity of $\sim$335 \scunit~utilizing one Na ion intercalation into every \ce{TiO2} formula unit.
Moreover, its electrode voltage appears to be as low as $\sim$0.56 V, indicating the increased energy density when used as anode material, and the result on ion migration also supports the applicability of this compound as SIB anode material.

\section{Methods}
\subsection{Structures of \ce{TiO2} polymorph with Na insertion}
\begin{figure}[!th]
\centering
\includegraphics[clip=true,scale=0.12]{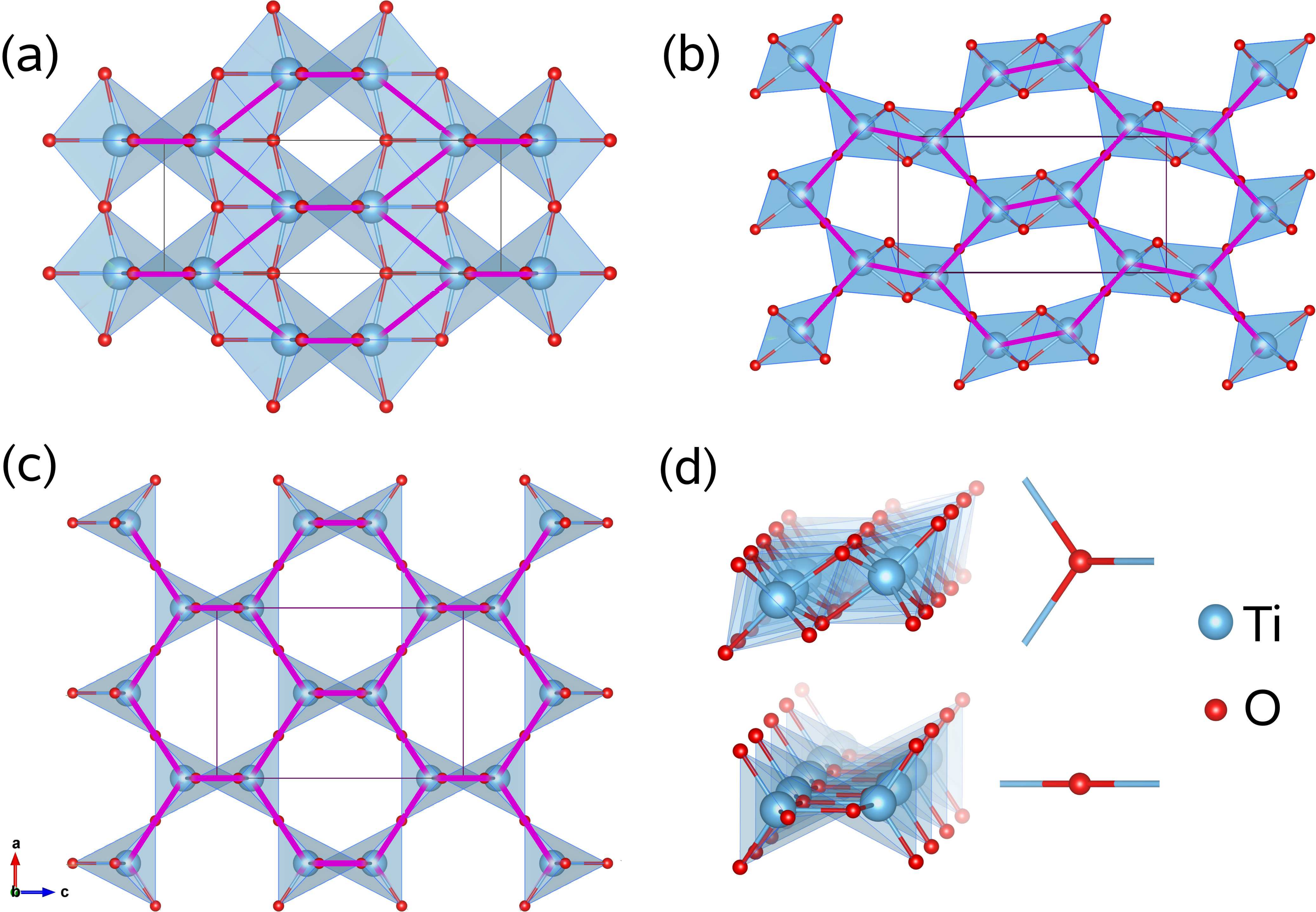}
\caption{\label{fig1}Polyhedral view of typical \ce{TiO2} polymorph in a) anatase, b) ramsdellite and c) trigonal bipyramid (TB) phases. d) Edge-sharing \ce{TiO6} octahedra (top) and edge- and corner-sharing \ce{TiO5} TB (bottom). Pink solid lines indicate the Ti hexagonal framework.}
\end{figure}

Titania has rich polymorph from traditional bulky phases of rutile, anatase and brookite to layered porous phases such as ramsdellite- and hollandite-like structures.
These titania phases can be characterized to be composed of \ce{TiO6} octahedra as their building block.
As newly suggested by Ma {\it et al.}~\cite{Sicong}, there is another building block of titania polymorph, namely penta-oxygen-coordinated \ce{TiO5} trigonal bipyramid (TB).
Figure~\ref{fig1} shows typically the anatase, ramsdellite and TB structures projected onto $a$-$c$ plane in polyhedral view.
As denoted by pink solid lines in this figure, these polymorph look very similar in respect of Ti$-$Ti linkage, which is a hexagonal framework, but they are different with respect to the building block and their interconnection.
In the anatase phase the \ce{TiO6} octahedra are connected by edge-sharing, forming the 3D channel inside the framework, while in the ramsdellite phase the octahedra are interconnected by edge- and corner-sharing, showing the tunnel-type 1D channel.
In these phases, the channels are too narrow for any guest ion to intercalate, because the oxygen atoms are placed deep inside the framework of Ti$-$Ti linkage.
In the TB structure, the \ce{TiO5} building blocks are interconnected by edge- and corner-sharing, forming the porous 1D channel with the oxygen atoms placed in-between Ti$-$Ti linkage when looked at in the direction of channel, which can offer sufficient space for ion host and migration.
It is worth noting that the pore size increases from anatase to ramsdellite and to TB phases.

There are four different crystalline phases from the TB building block~\cite{Sicong} (Figure S1), namely TB-I (orthorhombic, $Imma$), -II (orthorhombic, $Cmcm$), -III (tetragonal, $I_4/mmm$), and -IV (orthorhombic, $Imma$).
The TB-I and -II phases can be characterized to be composed of 6-member ring of \ce{TiO5}, while the TB-III and -IV phases to contain 8- and 4-member rings.
As mentioned above, they have a common character of 1D channel with the large pore sizes of 5.77 \AA~(TB-I), 5.65 \AA~(TB-II), 6.16 \AA~(TB-III) and 6.71 \AA~(TB-IV), which might be beneficial to the insertion and diffusion of alkali metal cations for the ion battery applications.
Contrary to our expectations, however, Li ions were found to place on the center of walls of the channel instead of the channel inside and diffuse across the channel rather than along the channel~\cite{Sicong}.

\begin{figure}[!th]
\centering
\includegraphics[clip=true,scale=0.06]{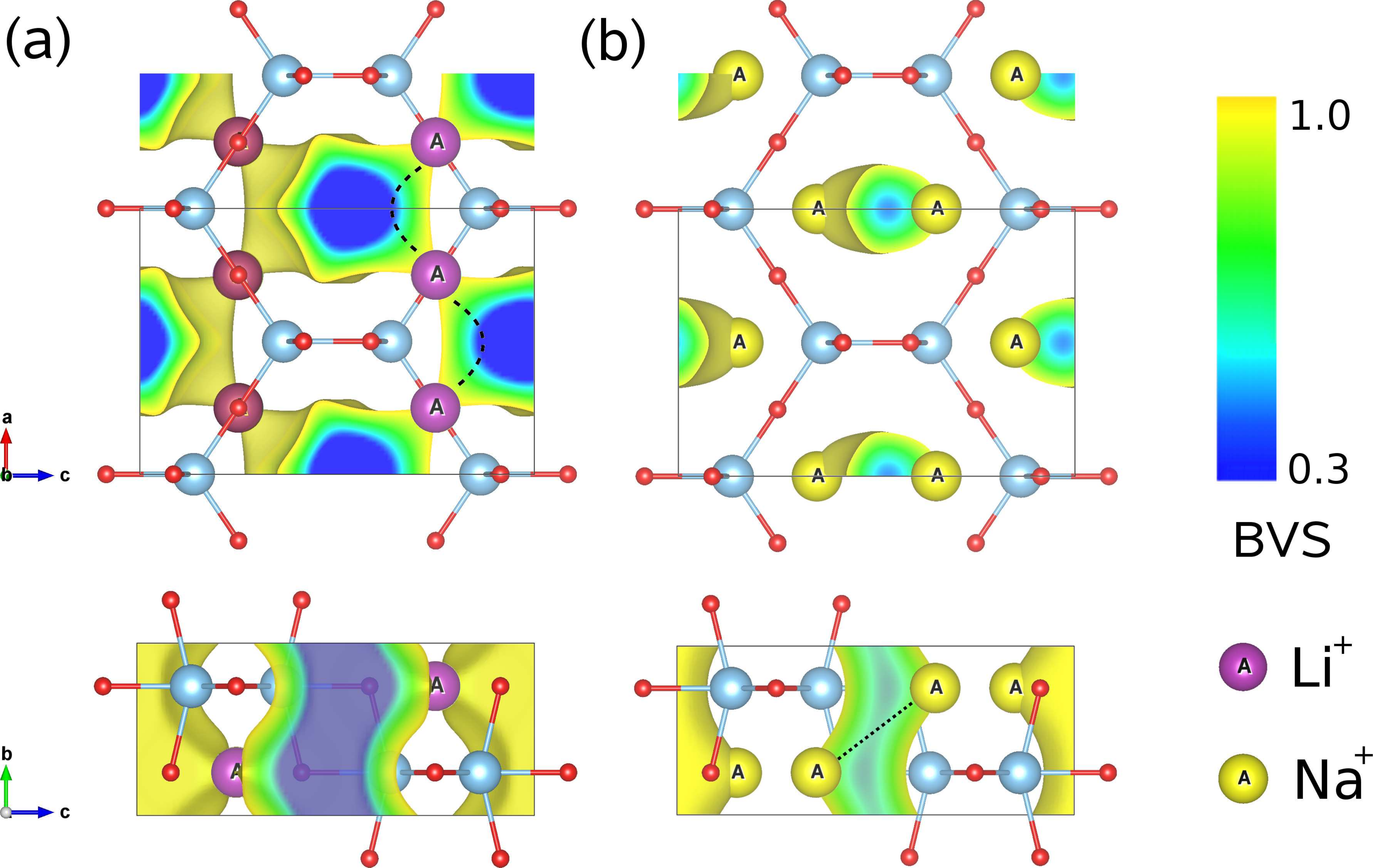}
\caption{\label{fig2}Isosurface view of bond valence sum (BVS) of \ce{TiO2} in TB-I structure, showing the anchoring sites of (a) Li and (b) Na ions, projected onto $a$-$c$ (top) and $b$-$c$ (bottom) planes. The identified insertion sites of Li and Na are illustrated as yellow- and pink-fillled circles with A label, and their migration paths are denoted as dashed lines.}
\end{figure}
Such abnormal Li insertion sites and diffusion pathways can be understood by estimating and analyzing the bond valence sum (BVS) in \ce{TiO2} bulk phases~\cite{Brown_BVS,Adam_BVS1,Adam_BVS3}.
Figure~\ref{fig2} shows the isosurface of BVS with the values from 0.3 to 1.0 in the crystalline \ce{TiO2} TB-I phase.
Since Li and Na ions can possess an oxidation state of +1 in compounds, they could be located on the isosurface of BVS = 1.
In accordance with Ref.~\cite{Sicong} and as shown in Figure~\ref{fig2}(a), Li ions with smaller ionic radius are verified to be located on the center of two bridge oxygen anions in the channel wall.
On the other hand, the insertion sites of Na ions with larger ionic radius are shown to be possibly inside the 1D channel and to migrate through the channel along the zigzag path by the BVS map for Na in Figure \ref{fig2}(b).
The BVS maps for Na in TB-II, -III and -IV are shown in Figure S1.
In the cases of TB-I and -II phases, two different Na sites are identified inside the 6-member ring channel, while in the cases of TB-III and -IV phases four different sites are possible inside the 8-member ring channel with the same zigzag migration paths.
Accordingly, these four different TB phases have a potential to hold one Na ion per \ce{TiO2} formula unit inside the channel, resulting in an exceptionally high capacity of $\sim$335 \scunit.
Such insertion sites and migration paths of Na ion roughly estimated from the BVS map should be rectified by DFT calculations.
The conventional unit cells of TB phases contain 4 \ce{TiO2} formula units in TB-I phase and 8 formula units in TB-II, -III and -IV phases, and thus 2$\times$1$\times$1 supercell for TB-I phase and the unit cells for other phases, containing the 8 formula units (24 atoms) consistently, were used in the DFT calculations.
In order to perform simulations of Na insertion and diffusion, we built the unit cells for Na-intercalated titania \ce{Na_{$x$}TiO2}, which contain Na ions from 1 to 8, corresponding to from $x=0.125$ to $x=1$.

\subsection{Computational details}
The pseudopotential plane-wave method within the DFT framework was used to carry out the first-principles calculations, as implemented in {\small Quantum ESPRESSO} (QE) package (version 6.2)~\cite{QE}.
The project-augmented-wave (PAW) pseudopotentials from the pslibrary~\footnote{We use the PAW pseudopotentials Na.pbesol-spn-kjpaw\_psl.1.0.0.UPF, Ti.pbesol-spn-kjpaw\_psl.1.0.0.UPF and O.pbesol-nl-kjpaw\_psl.1.0.0.UPF from http://pslibrary.quantum-espresso.org.} were used to describe the ion-electron interaction, and the revised version for solid of Perdew-Burke-Ernzerhof functional (PBEsol) within the generalized gradient approximation (GGA)~\cite{PBEsol} was used to describe the exchange-correlation (XC) interaction between the valence electrons.
The wave functions of valence electrons and electronic densities were expanded by using the plane wave basis sets generated with the cut-off energies of 70 and 800 Ry, respectively.
Special $k$-points were set to be (2$\times$6$\times$2) for structural optimization.
These computational parameters guarantee a total energy accuracy of 5 meV per formula unit. The positions of all atoms and lattice parameters were fully relaxed until the atomic forces converge to 0.02 eV \AA$^{-1}$.
For the calculations of density of states (DOS) and charge difference, denser $k$-points of (4$\times$6$\times$4) were used.

The migration paths and the corresponding activation energies for sodium ion migrations were determined by using the climbing image nudged elastic band (NEB) method~\cite{NEB} as implemented in the neb.x code of the QE package.
During the NEB run, the supercell lattice parameters were fixed at the optimized values, while all the atomic positions were allowed to relax.
The seven NEB image points were used, and the convergence criteria for the force orthogonal to the path was set to be 0.05 eV \AA$^{-1}$.

\section{Results and Discussion}
\subsection{Preliminary check of XC functional}
We first checked a reliability of the selected PBEsol XC functional with the crystalline lattice constants of the precedent rutile, anatase, hollandite and ramsdellite \ce{TiO2} phases determined by structural optimizations in comparison with those by PBE calculation and experiment.
Table~\ref{table1} presents their lattice constants calculated by using the PBEsol functional in this work and the PBE functional, together with the experimental values.
The PBEsol functional was confirmed to reproduce the lattice constants of these phases with the relative error of under $\pm$1\% in comparion with experiment, being clearly better than the PBE functional which overall overestimated the lattice constants in accordance with the general tendency of GGA for XC interaction.
In the following calculations, therefore, we could use the PBEsol functional with a confidence of accuracy.
\begin{table}[!th]
\small
\caption{\label{table1}Lattice constants of \ce{TiO2} crystalline solids with rutile, anatase, hollandite and ramsdellite phases, calculated by using the PBEsol functional in this work.} 
\begin{tabular}{@{\hspace{2pt}}l@{}c@{\hspace{4pt}}c@{\hspace{4pt}}c@{\hspace{4pt}}c@{\hspace{4pt}}c@{\hspace{4pt}}c@{\hspace{4pt}}c@{\hspace{2pt}}}
\hline 
 & \multicolumn{3}{c}{$a = b$ (\AA)} && \multicolumn{3}{c}{$c$ (\AA)} \\
\cline{2-4} \cline{6-8}
\footnotesize Polymorph & \footnotesize PBEsol & \footnotesize PBE$^ a$ & \footnotesize Exp.$^b$ && \footnotesize PBEsol & \footnotesize PBE$^ a$ & \footnotesize Exp.$^b$ \\
\hline
\footnotesize Rutile    & 4.600  & 4.66  & 4.593  && 2.939 & 2.97 & 2.959 \\
\footnotesize Anatase        & 3.779  & 3.82  & 3.785  && 9.511 & 9.70 & 9.514 \\
\footnotesize Hollandite     & 10.184 & 10.31 & 10.161 && 2.951 & 2.98 & 2.910 \\
\footnotesize Ramsdelite$^c$ & 4.865  & 5.06  & 4.902  && 2.951 & 3.16 & 2.958 \\
               & 9.462  & 9.60  & 9.459  &&       &      & \\       
\hline
\multicolumn{8}{l}{{\scriptsize $^a$ Results from Ref.~\cite{Sicong}}} \\
\multicolumn{8}{l}{{\scriptsize $^b$ Rutile, anatase and ramsdellite from Ref.~\cite{ICSD} and hollandite from Ref.~\cite{Latroche}}} \\
\multicolumn{8}{l}{{\scriptsize $^c$ In the ramsdellite phase, $a\neq b$, and $b$ is shown below line.}}
\end{tabular}
%\flushleft \scriptsize
%$^a$ Ref.~\cite{Sicong} \\
%$^b$ Rutile, anatase and ramsdellite from Ref.~\cite{ICSD} and hollandite from Ref.~\cite{Latroche} \\
%$^c$ In ramsdellite phase, $a\neq b$, and $b$ is shown below line.
\end{table}
\begin{figure}[!b]
\centering
\includegraphics[clip=true,scale=0.65]{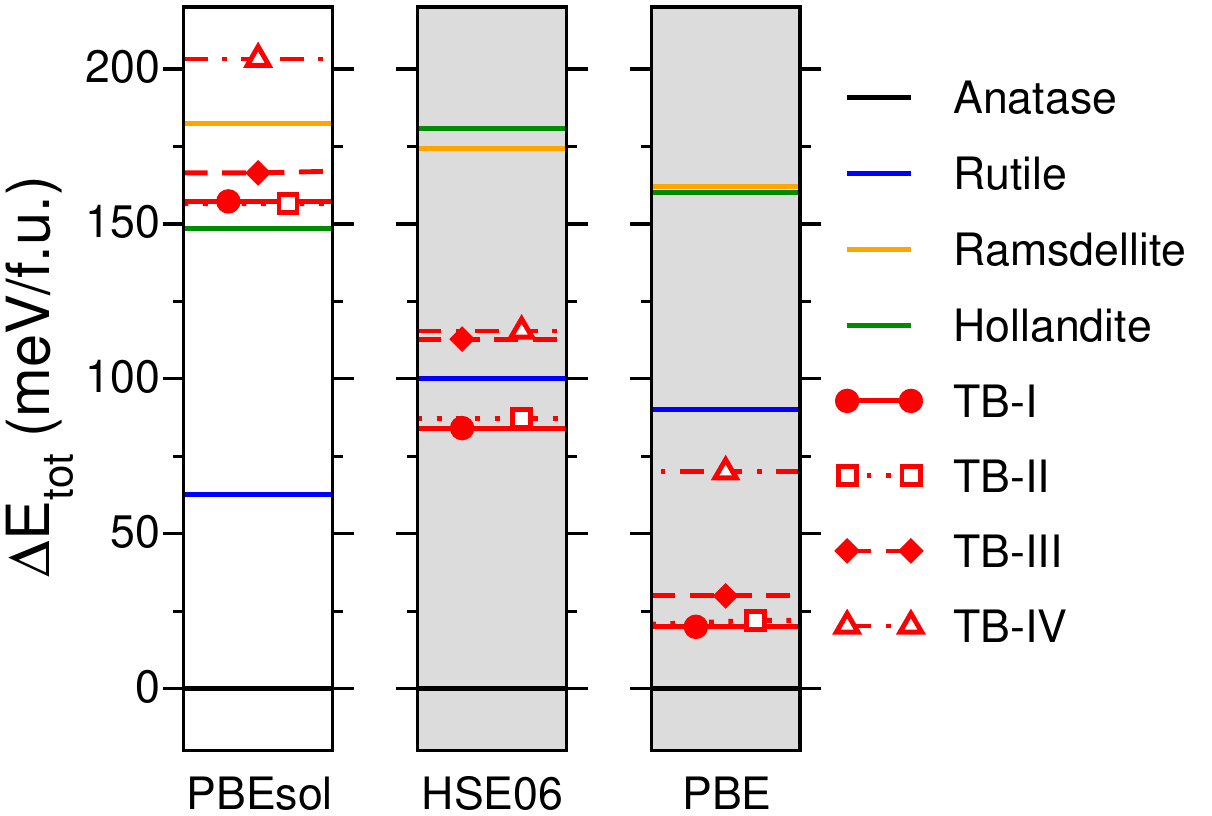}
\caption{\label{fig3}Schematic view of relative stabilities of \ce{TiO2} polymorph with the total energy difference per formula unit in reference to the anatase phase, calculated by using the PBEsol functional. Those obtained by using the HSE06 and PBE functionals in the previous work~\cite{Sicong} are shown in left panel with grey background.}
\end{figure}
We then investigated the relative stabilities of  the \ce{TiO2} polymorph by comparing their calculated DFT total energies per formula unit after structural optimizations for anatase, rutile, ramsdellite, hollandite, and TB-I, -II, -III and -IV phases.
Figure~\ref{fig3} shows the schematic view of total energy differences of \ce{TiO2} phases in reference to the anatase phase, together with the previous DFT results obtained by using the PBE and the hybrid HSE06 functionals~\cite{Sicong}.
In these DFT calculations, the anatase phase was confirmed to have the lowest total energy.
One can perceive that the overall tendency of the phases' stabilities changes severely depending on the choice of XC functional.
In the previous PBE and HSE06 calculations, TB phases were found to be energetically more stable than the precedent phases of rutile, ramsdellite and hollandite (the TB-III and -IV phases were $\sim$6 meV higher per formula unit than the rutile phase).
In our PBEsol calculation, however, TB-I, -II, and -III phases were found to be energetically higher than the anatase, rutile and hollandite phases and lower than the ramsdellite phase, while TB-IV phase to have the highest total energy among the phases.
In addition, TB-I and -II phases had almost the same total energies commonly with the three different XC functionals, indicating a close similarity of their structures.
It should be noted that the rational phase stability diagram as a function of temperature and pressure can be determined by estimating the Gibbs free energy with the first-principles atomistic thermodynamics method.

\subsection{Energetics and structural properties}
In order to estimate the cycling stability of electrode materials during insertion/extraction of Na ions, we investigated the crystalline structures of Na-inserted \ce{TiO2} compounds \ce{Na_{$x$}TiO2} in TB phases, and calculated the formation energies of these compounds and the binding energies of Na ion.
As mentioned above, our supercells for TB phases include 8 formula units and thus the number of inserted Na ions could be from 0 to 8, allowing the Na content $x$ = 0.0, 0.125, 0.25, 0.375, 0.5, 0.625, 0.75, 0.875, and 1.0 in \ce{Na_{$x$}TiO2}.
In Table~\ref{table2}, we show the calculated lattice constants ($a$, $b$, $c$), relative volume expansion rate $r_{\text{vol}}=(V_x-V_0)/V_0\times100$\% as increasing the Na content $x$ in \ce{Na_{$x$}TiO2} compounds in TB-I and -II phases, and binding energies of Na ion ($E_{\text{b}}$) (Those for TB-III and -IV phases are given in Table S1.)
\begin{table}[!th]
\small
\caption{\label{table2}Lattice constants $a$, $b$, and $c$ (\AA), relative volume expansion rate $r_{\text{vol}}=(V_x-V_0)/V_0\times100$\%, and sodium binding energy $E_{\text{b}}$ (eV), in \ce{Na_{$x$}TiO2} compounds with TB-I and -II phases, calculated with the PBEsol XC functional.}
\begin{tabular}{l@{\hspace{10pt}}c@{\hspace{10pt}}c@{\hspace{10pt}}c@{\hspace{10pt}}c@{\hspace{10pt}}c}
\hline
$x$ & $a$ & $b$ & $c$ & $r_\text{vol}$ & $E_\ce{b}$ \\
\hline
& \multicolumn{5}{l}{\ce{Na_{$x$}TiO2} (TB-I)} \\
0.0  & 11.963  & 3.771  & 8.886  & $-$   &  $-$       \\
1/8  & 11.978  & 3.812  & 8.831  & 0.58  & $-$2.20  \\
2/8  & 11.820  & 3.821  & 9.005  & 1.46  & $-$2.13  \\
3/8  & 11.852  & 3.907  & 8.853  & 2.26  & $-$1.94  \\
4/8  & 11.506  & 3.920  & 9.161  & 3.08  & $-$1.84  \\
5/8  & 11.219  & 3.923  & 9.377  & 2.96  & $-$1.79  \\
6/8  & 11.021  & 3.936  & 9.539  & 3.23  & $-$1.77  \\
7/8  & 10.879  & 3.945  & 9.648  & 3.28  & $-$1.74  \\
1.0  & 10.763  & 3.956  & 9.742  & 3.48  & $-$1.73  \\
\hline
& \multicolumn{5}{l}{\ce{Na_{$x$}TiO2} (TB-II)} \\
0.0  & 10.697  & 3.774  & 9.926  & $-$     & $-$     \\
1/8  & 10.582  & 3.798  & 9.543  & $-$4.28 & $-$2.29 \\
2/8  & 10.505  & 3.811  & 9.227  & $-$7.82 & $-$2.34 \\
3/8  & 10.225  & 3.874  & 9.520  & $-$5.88 & $-$2.04 \\
4/8  & 10.055  & 3.949  & 9.760  & $-$3.27 & $-$1.93 \\
5/8  & 10.000  & 4.010  & 9.680  & $-$3.13 & $-$1.76 \\
6/8  & 10.010  & 3.990  & 9.821  & $-$2.10 & $-$1.63 \\
7/8  & 10.134  & 3.974  & 9.805  & $-$1.46 & $-$1.50 \\
1.0  & 10.086  & 4.012  & 9.814  & $-$0.89 & $-$1.48 \\
\hline
\end{tabular}
\end{table}
\begin{figure*}[!th]
\centering
\begin{tabular}{ccc}
\includegraphics[clip=true,scale=0.14]{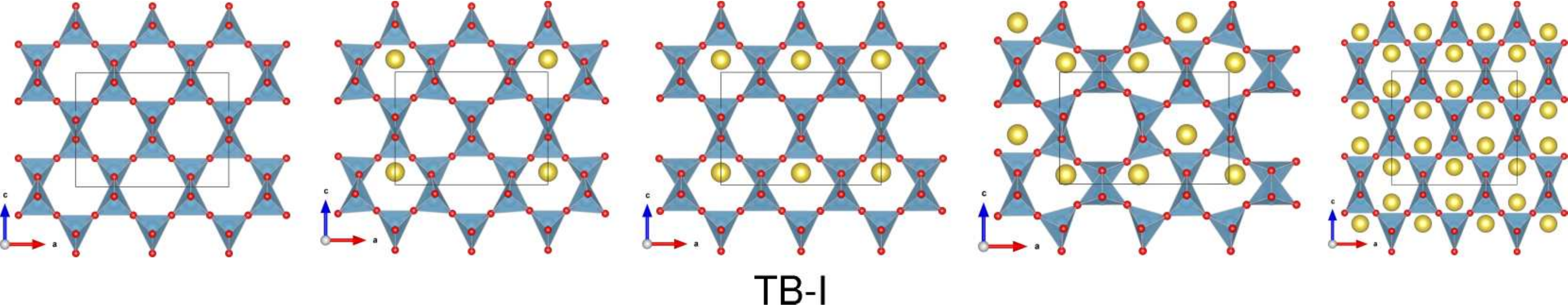} & &
\includegraphics[clip=true,scale=0.14]{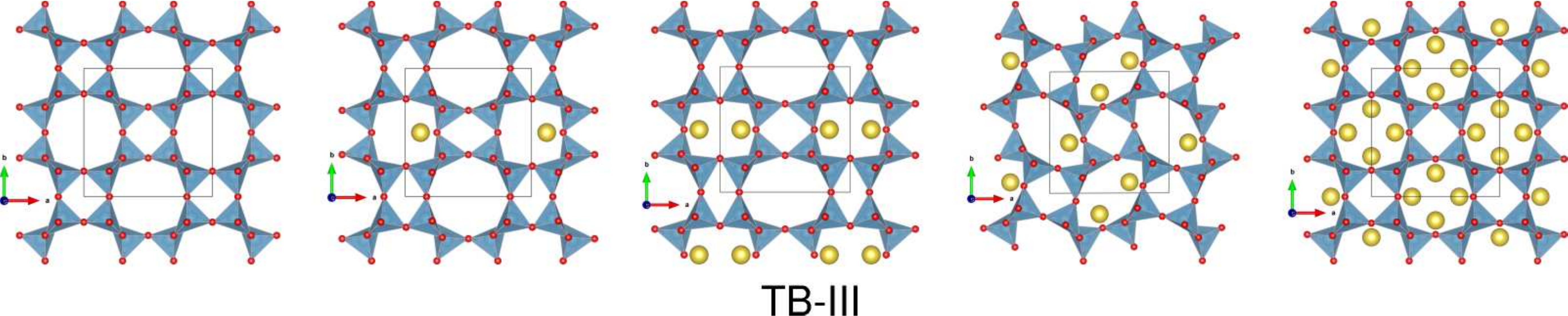} \\
\includegraphics[clip=true,scale=0.5]{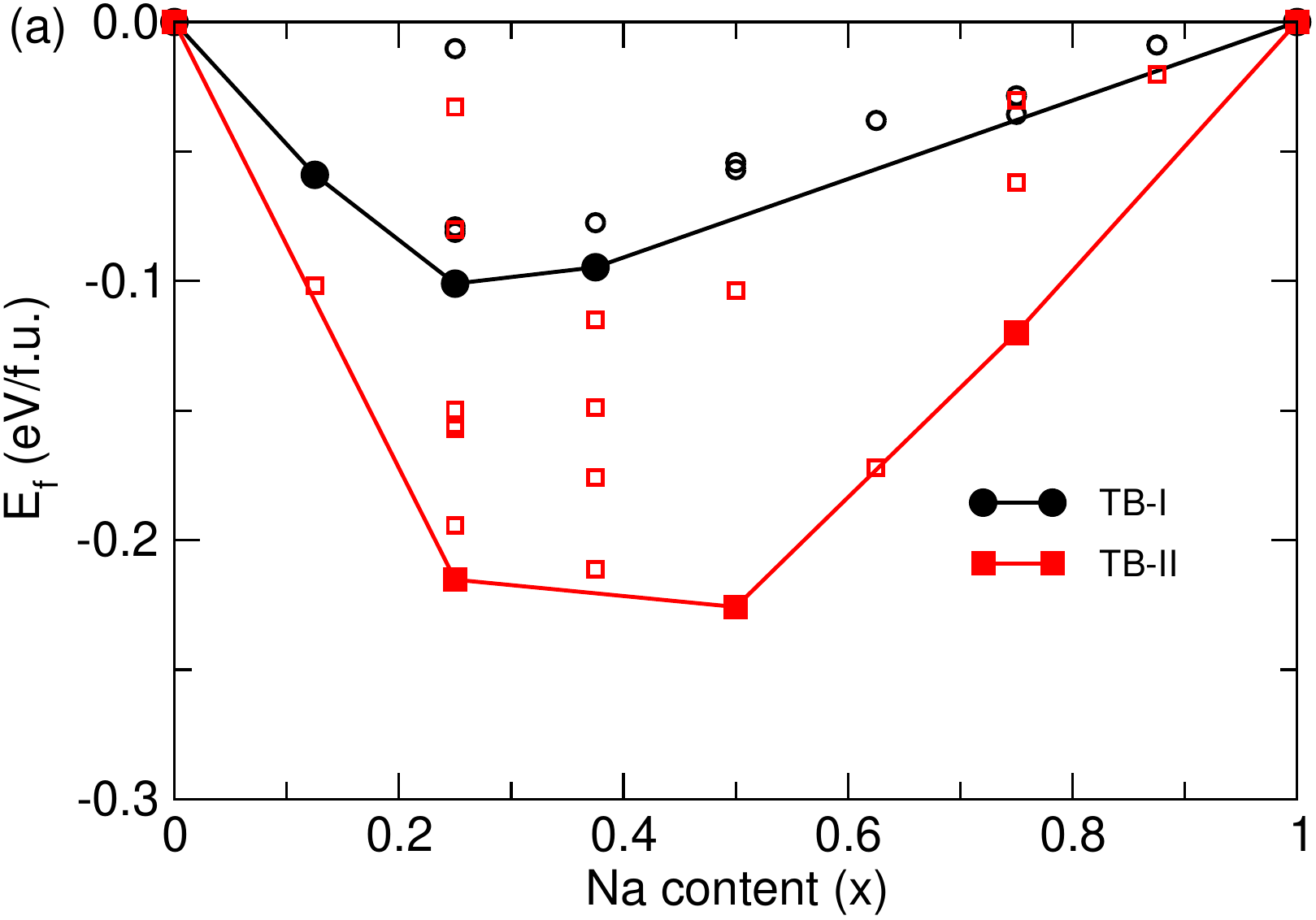} & &
\includegraphics[clip=true,scale=0.5]{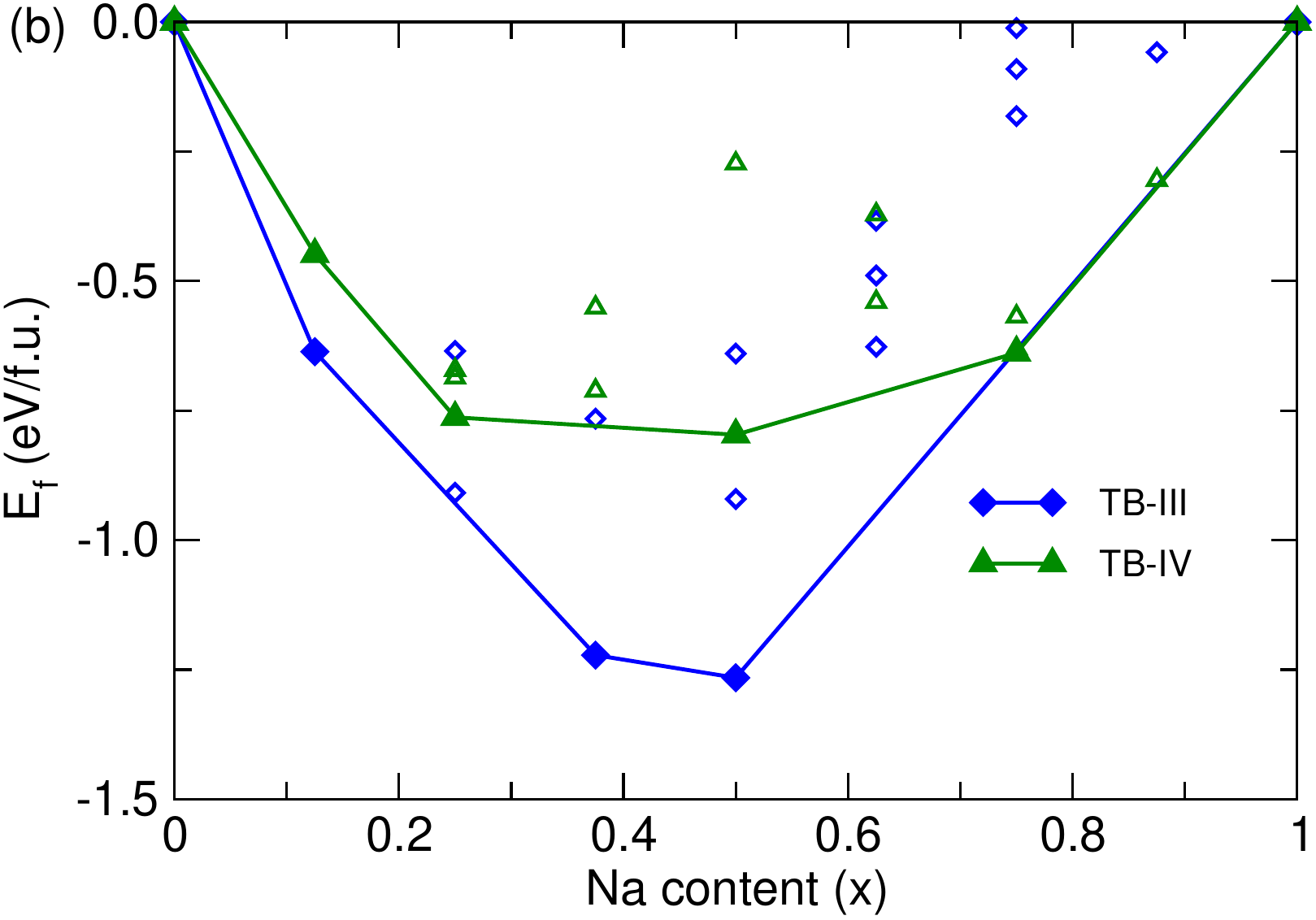} \\
\includegraphics[clip=true,scale=0.14]{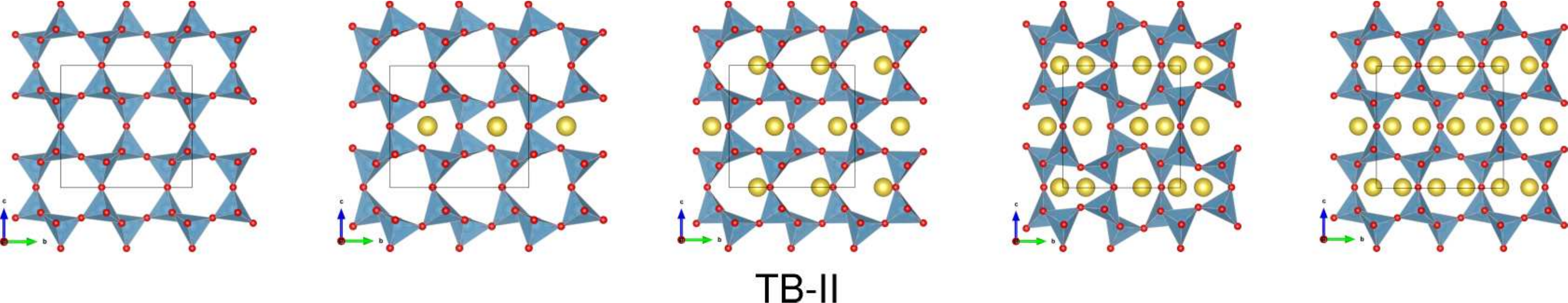} & &
\includegraphics[clip=true,scale=0.14]{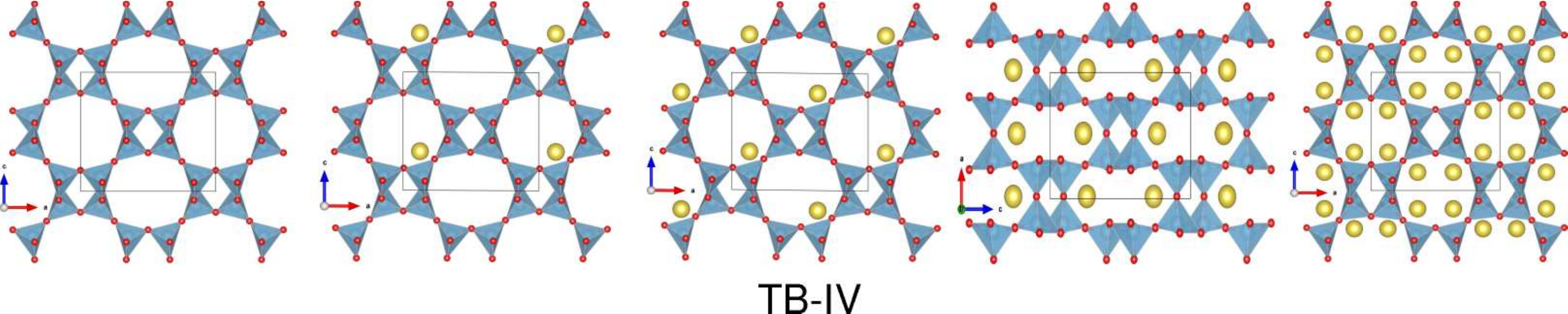} \\
\end{tabular}
\caption{\label{fig4}Convex hull plot of formation energy per formula unit (f.u.) of \ce{Na_{$x$}TiO2} compounds in (a) TB-I and -II phases and (b) TB-III and -IV phases with Na content $0\leq x\leq1$, calculated by using the PBEsol XC functional. The optimized supercells corresponding to the points lying on the convex hull line are shown together.}
\end{figure*}

The strength of the interaction between Na atom and the host compound \ce{TiO2} can be assessed by the binding energy per Na atom ($E_{\text{b}}$), which can be defined as follows,
\begin{equation}
E_{\text{b}}=\frac{1}{x}\left[E_{\ce{Na}_x\ce{TiO2}} - \left(E_\ce{TiO2} + xE_\ce{Na_{gas}}\right)\right] \label{eq_ebind}
\end{equation}
where $E_\ce{Na_{x}TiO2}$ and $E_\ce{TiO2}$ are the total energies per formula unit of \ce{Na_{$x$}TiO2} and \ce{TiO2} compounds, respectively, and $E_\ce{Na_{gas}}$ is the total energy of Na in gas phase ({\it i.e.}, isolated Na atom).
The sodium binding energies were calculated to be negative in the four different TB phases, giving an implication of exothermic process for Na insertion, and they decrease in magnitude as increasing the Na content $x$ in overall (see Figure S3).
Looking at the TB-I case carefully, $E_{\text{b}}$ gradually decreases in magnitude from $-$2.20 eV ($x$ = 0.125) to $-$1.84 eV ($x$ = 0.5) but changes a little only from $-$1.79 eV ($x$ = 0.625) to $-$1.73 eV ($x$ = 1.0) when more than 4 Na atoms are inserted into the supercell.
As expected from the above analysis of BVS in Figure~\ref{fig2}(b), the Na atoms were confirmed to reside at one of two identical sites inside the 6-member ring by structural optimizations of \ce{Na_{$x$}TiO2} supercells.
Here, the Na atoms might be supposed to place uniformly over the rings first, so that each Na atom can occupy the whole channel space when $x \leq 0.5$, resulting in the growing repulsion between the inter-channel Na atoms and thus decreasing the binding energy in magnitude.
When Na atoms are further inserted into the \ce{Na_{0.5}TiO2} compound, they could occupy another remaining site inside the ring, and could be forced to move towards the 1D channel wall owing to the repulsion between the intra-channel Na atoms, leading to a strengthening of Na$-$O attraction and very small change of $E_{\text{b}}$ when $x > 0.5$.
We note that the similar trend and analysis can also be valid for TB-III and -IV phases, where the binding energies were found to change within $-$2.21 $\sim$ $-$1.89 eV ($-$1.70 $\sim$ $-$1.58 eV) and $-$2.10 $\sim$ $-$1.86 eV ($-$1.76 $\sim$ $-$1.66 eV) in the range of 0.125 $<x<$ 0.5 (0.625 $<x<$ 1.0), respectively.
In TB-II, however, $E_b$ decreases in magnitude from $-$2.29 to $-$1.93 eV when $x \leq 0.5$ and from $-1.76$ to $-1.48$ eV when $x > 0.5$, which might be a relatively larger change compared with the former cases.

During the charge/discharge process, electrodes in general experience a multi-step phase transition, induced by inserting the guest ions, and the relative phase stability of the intermediate compound with a certain ratio $x$ of inserted ion can be estimated by the formation energy.
The formation energy of Na-inserted \ce{TiO2} compound, \ce{Na_{$x$}TiO2}, from the two end compounds, {\it i.e.}, pristine \ce{TiO2} and fully sodiated \ce{NaTiO2}, can be calculated using their corresponding total energies as follows,
\begin{equation}
E_{\text{f}}=E_{\ce{Na}_x\ce{TiO2}}-\left[xE_{\ce{NaTiO2}}+(1-x)E_{\ce{TiO2}}\right] \label{eq_eform}
\end{equation}
The calculated formation energies of \ce{Na_{$x$}TiO2} (0 $\leq x \leq$ 1) compounds in the four different TB phases are shown in Figure~\ref{fig4}, together with the convex hull lines.
For all the intermediate compounds \ce{Na_{$x$}TiO2} in these TB phases, the formation energies were determined to be negative, indicating their certain formation from the two end compounds.
The following observations were obtained; (1) the lowest energy compounds are found at $x=0.25$ in TB-I and -III phases while at $x=0.5$ in II and IV phases, (2) the magnitudes of the lowest energies are in the order of IV ($-$1.28 eV) $>$ III ($-$1.26) $>$ II ($-$0.23) $>$ I phase ($-$0.10), (3) the intermediate phases lying on the convex hull lines are found at $x=0.125, 0.25, 0.375$ in TB-I phase, $0.25, 0.5, 0.75$ in II phase, $0.125, 0.25, 0.375$ in III phase and $0.125, 0.25, 0.5$ in IV phase.
These intermediate phases can be commonly characterized to preserve their original crystalline lattices of the pristine \ce{TiO2} TB phases as orthorhombic systems, {\it i.e.}, the lattice angles as $\alpha = \beta = \gamma = 90^\circ$, but with different lattice constants and slightly distorted \ce{TiO5} trigonal bipyramids and their interconnections of forming the 6- or 8-member rings, and will be used to evaluate the electode step potential as discussed below.
It should be noted that although other intermediate phases with different Bravais lattice could be generated also during the insertion process, they were drifted away from the convex hull lines.
These indicate that severe phase transformations in these compounds might not occur during the charge/discharge process, giving a hint of good cyclic stability if they would be utilized as electrode materials.

In order to get more detailed insight of cyclic stability, we considered the change tendency of lattice constants and the relative volume expansion rate ($r_{\text{vol}}$) as increasing the Na content $x$.
The change tendencies of lattice constants and volume were observed to be quite different according to the phase.
In TB-I phase, the lattice constant $a$ gradually decreases from 11.96 to 10.76 \AA~as increasing $x$, while $b$ and $c$ increase from 3.77 and 8.89 \AA~to 3.96 and 9.74 \AA, resulting in the gradual increase of $r_{\text{vol}}$ from 0.58 to 3.48\% (see Table~\ref{table1} and Table S1).
Similar tendency was observed in TB-IV phase, where $r_{\text{vol}}$ increases from 0.48 to 3.73\% as increasing $x$ with the lowest value of 0.33\%~at $x=0.5$.
For the case of TB-II phase, the lattice constants $a$ and $c$ decrease in general while $b$ increase as increasing $x$, leading to the volume shrink to $-$7.82\%~at $x=0.25$ and again the expansion to $-$0.89\%~at $x=1.0$ but still shrink compared with that of $x=0.0$.
Similar wavy change of lattice constants were also found in TB-III, where $a$ and $c$ decrease first and then increase, resulting in the decrease of $r_{\text{vol}}$ to $-$4.16\%~at $x=0.25$ and then the increase to 9.10\%~at $x=1.0$, as increasing $x$.
With respect to the battery performance, therefore, the TB-I and -IV phases are favorable due to relatively lower volume expansion rate than II and III phases.
It should be again emphasized that in these TB phases the relative volume change is under 10\%, and moreover, there is no change of crystalline system, indicating their higher cycling stabilities during the charge/discharge process.

\subsection{Electrochemical properties}
We then estimated the electrochemical performance of \ce{Na_{$x$}TiO2} compounds in the TB phases by calculating the electrode potential and the activation energy for sodium ion migration.
To compute the step intercalation electrode potential {\it vs}. Na/\ce{Na+} within the DFT framework, we considered the following general cell reaction,
\begin{equation*}
(x_j-x_i)\ce{Na+}+(x_j-x_i)e^-+\ce{Na}_{x_i}\ce{TiO2} \rightarrow \ce{Na}_{x_j}\ce{TiO2}
\end{equation*}
where $x_j$ and $x_i$ are the adjacent Na contents, which can be identified in the convex hull plot of formation energy, and $e$ is the elementary charge.
Then, the step discharge voltage can be calculated by taking the DFT total energy difference in between the adjacent Na-inserted \ce{TiO2} compounds, by using the following equation~\cite{Aydinol97prb},
\begin{equation}
V = -\frac{E_{\ce{Na}_{x_j}\ce{TiO2}}-E_{\ce{Na}_{x_i}\ce{TiO2}}-(x_j-x_i)E_{\ce{Na_{bcc}}}}{e(x_j-x_i)} \label{eq_voltage}
\end{equation}
where $E$ is the DFT total energy of the corresponding compound.

\begin{figure}[!th]
\centering
\includegraphics[clip=true,scale=0.5]{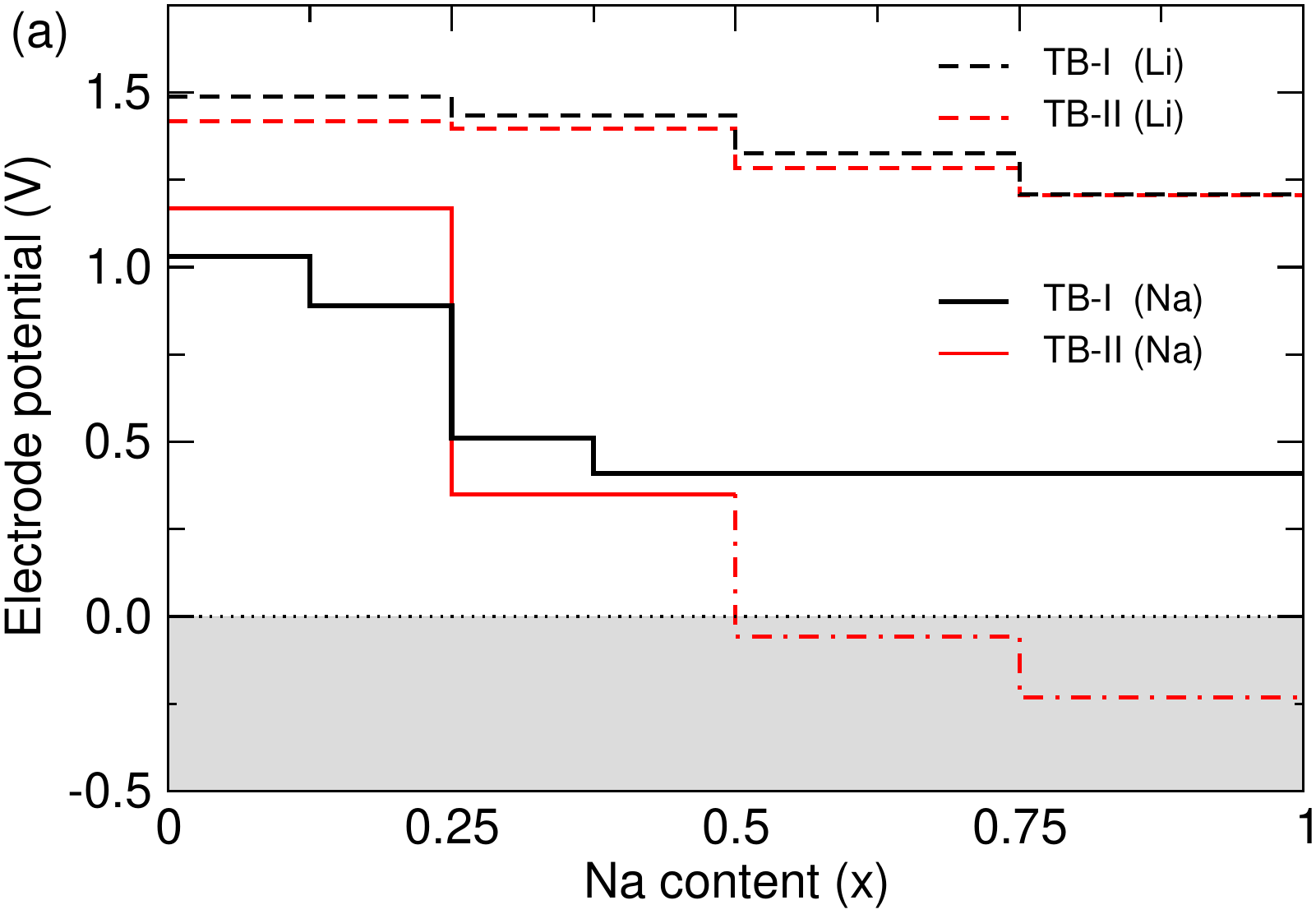}
\includegraphics[clip=true,scale=0.5]{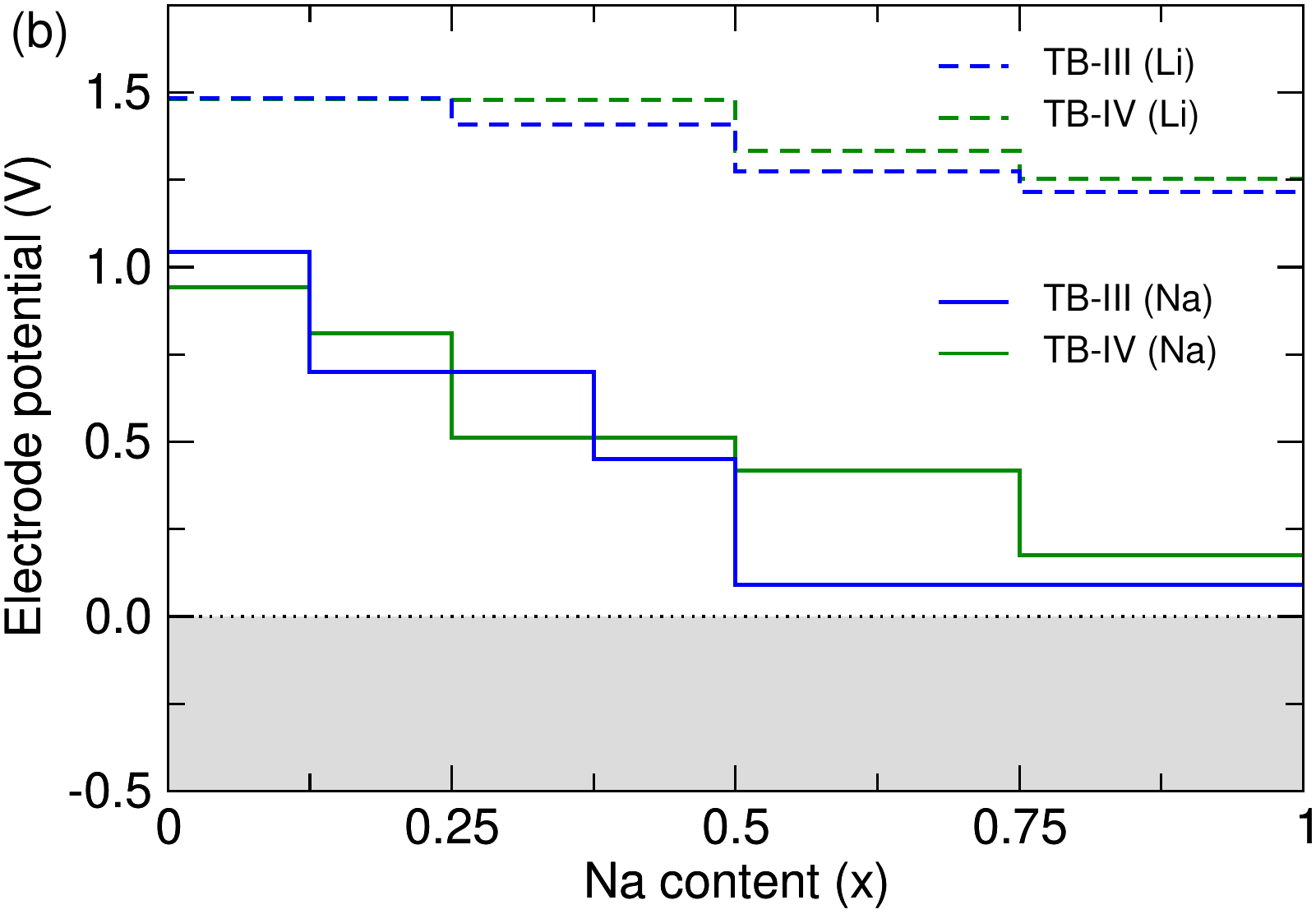}
\caption{\label{fig5}Electrode potential {\it vs.} Na/\ce{Na+} in \ce{Na_{$x$}TiO2} compounds with (a) TB-I and -II phases and (b) -III and -IV phases, calculated using the intermediate structures found on the convex hull shown in Figure~\ref{fig4}. Those for Li intercalation are shown as reference~\cite{Sicong}.}
\end{figure}
Figure~\ref{fig5} displays the average redox potential {\it vs.} Na/\ce{Na+}, calculated using the intermediate structures found on the convex hull line (Figure~\ref{fig4}).
The previously calculated electrode voltages of TB compounds as LIB anodes~\cite{Sicong} are also shown for comparison, demonstrating that the average voltages for SIB are clearly lower than their LIB counterparts as in many other electrode materials due to lower redox potential of Na than Li~\cite{Ong}.
As discussed above, the different Na insertion sites from the Li sites may also affect the electrode voltage, lowering it possibly due to less closely packed Na atoms with O atoms.
In the first step of the voltage profile, the redox potentials were computed to be 1.03 and 0.94 V between $x=0$ and $x=0.125$ in TB-I and -IV phases, and 1.17 and 1.04 V between $x=0$ and $x=0.25$ in TB-II and -III phases, respectively.
When compared with other Ti-O system, these values of voltage are higher than 0.7$\sim$0.8 V in \ce{Na_{2+$x$}Ti6O13}~\cite{Shen14}.
Increasing the Na content lowers the electrode voltage, and the final step voltages were calculated to be 0.41, 0.16, and 0.17 V from $x=0.375, 0.375$, and 0.5 to $x=1.0$ in the TB-I, -III, and -IV phases, respectively.
In the case of TB-II phase, the voltage becomes negative from $x=0.5$, implying that Na ions are energetically unfavorable to intercalate into the partially sodiated \ce{Na_{0.5}TiO2} compound.
These indicate that TB-I, -III, and -IV phases can support the fully sodiated compounds, reaching the highest specific capacity of 335 \scunit, whereas TB-II phase cannot be fully sodiated by electrochemical process, thus exhibiting the lower specific capacity.
\begin{figure*}[!th]
\centering
\includegraphics[clip=true,scale=0.14]{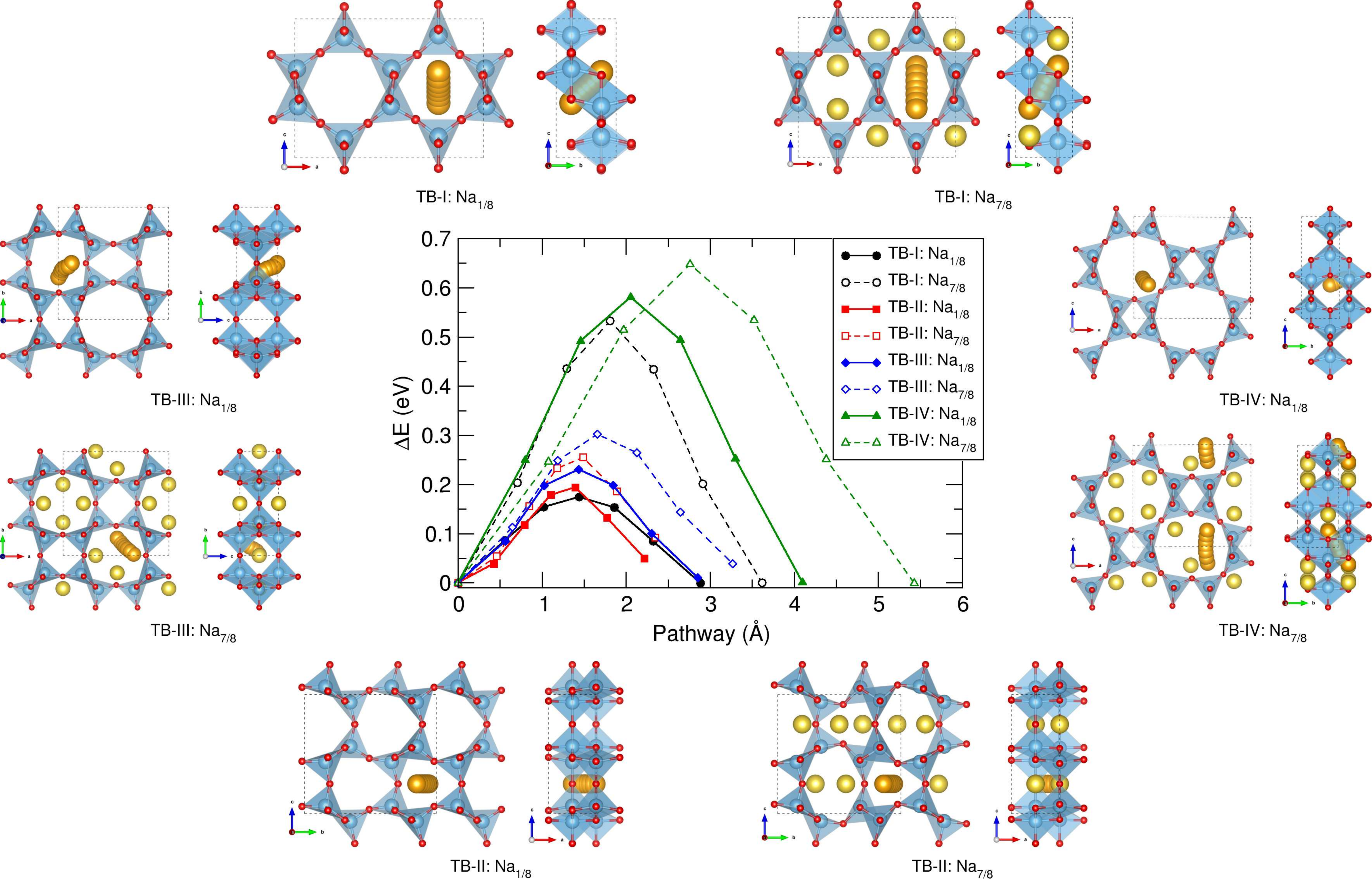}
\caption{\label{fig6}Energy profile for the Na ion migrations in \ce{Na_{$x$}TiO2} compounds with $x=0.125$ and 0.875 with four different TB phases, calculated with the PBEsol XC functional and NEB method.}
\end{figure*}
Therefore, regarding the electrode voltage, it is revealed that TB-I, -III, and -IV phases are favorable to SIB anode application, whereas TB-II phase may have relatively low specific capacity and exhibit serious safety issue of dendrite growth.

In order to assess the rate capability of TB compounds, which is one of the important factors for evaluating the suitability of electrode materials, we calculated the activation energies for Na ion migrations by using the NEB method.
Here, we considered two different limit conditions of Na concentraion, {\it i.e.}, Na-poor ($x=0.125$) and Na-rich ($x=0.875$) conditions.
Figure~\ref{fig6} shows the energy profile for Na ion migrations along the zigzag path through the 1D channel identified by the BVS analysis, and Table~\ref{table3} lists the corresponding activation barriers.
At the Na-poor condition, the activation barriers in TB-I, -II, and -III phases were estimated to be 0.17, 0.19, and 0.23 eV with migration distances of 2.8, 2.3, and 2.8 \AA, respectively, being much lower than $\sim$0.4 eV for Li ion migration in graphite as the standard LIB anode~\cite{Nobuhara} and for Na(digl)$_2$ solvated ion migration in graphite~\cite{yucj14}.
It was determined to be 0.58 eV with a migration distance of 4.1 \AA~in the case of TB-IV phase.
Considering that the pore sizes of 1D channel are 5.77, 5.65, 6.16, and 6.71 \AA~in TB-I, -II, -III, and -IV phases, respectively~\cite{Sicong}, it can be said that larger pore size leads to higher activation barrier.
When increasing the Na content, the activation barriers became higher as 0.53, 0.26, 0.30, and 0.65 eV in these TB phases at the Na-rich condition (Table~\ref{table3}).
In the case of TB-II phase, the Na-intercalated compound can be formed only up to \ce{Na_{0.5}TiO2}, revealed by the electrode potential, and thus the activation barrier at $x=0.5$ was also calculated to be 0.46 eV, indicating a lowering of ionic mobility (see Table S2).
Meanwhile, TB-I phase retains its superior rate capability up to $x=0.5$; the activation barriers were found to be 0.23, 0.21 and 0.34 eV at $x=0.25, 0.375$ and 0.5, respectively, being comparable with that for Na diffusion in layered dichalcogenides like \ce{MoS2} (0.28 eV)~\cite{Mortazavi14jps}.
Therefore, TB-I phase can be said to be the best choice for SIB anode application with respect to the reaction kinetics as well as volume change and electrode potential.
Even though TB-IV phase may not be much suitable for a fast charge/discharge kinetics, it is still worth considering as SIB anode material with its moderate ionic mobility.
\begin{table}[!th]
\small
\caption{\label{table3}Activation energies ($E_{\text{a}}$) for Na ion migrations in \ce{Na_{$x$}TiO2} compounds in TB phases at $x=0.125$ and 0.875.} 
\begin{tabular}{lcccc}
\hline
 & \multicolumn{4}{c}{$E_{\text{a}}$ (eV)} \\
\cline{2-5}
Na content & TB-I & TB-II & TB-III & TB-IV \\
\hline
0.125   & 0.17 & 0.19  & 0.23   & 0.58  \\
0.875   & 0.53 & 0.26  & 0.30   & 0.65  \\
\hline
\end{tabular}
\end{table}

\subsection{Electronic structures}
Finally, we calculated the electronic structures of the Na-intercalated \ce{TiO2} compounds in TB phases to get an insightful understanding of electron transfer and chemical bond.
From the viewpoint of battery operation, the electrode materials should be both electron and ion conductors, because the electrons traveled through the external circuit merge with the \ce{Na+} ions in the electrode passing through it.
Based on the above consequence that TB-I phase has the most superior electrochemical properties among the four TB phases, we only considered the TB-I phase in this subsection by calculating the projected density of states (PDOS) and electronic charge density differences.

\begin{figure}[!th]
\centering
\includegraphics[clip=true,scale=0.65]{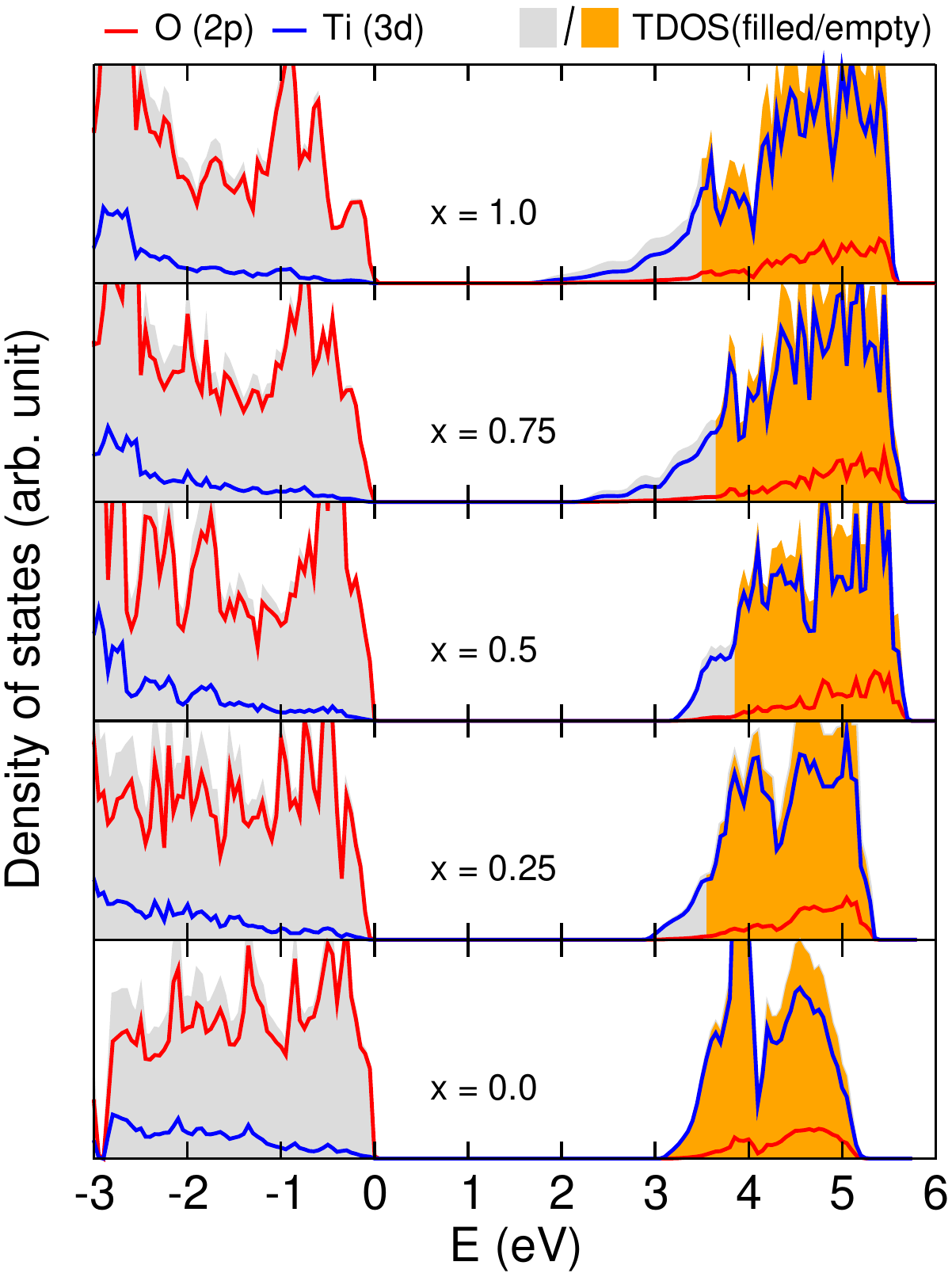}
\caption{\label{fig7}Projected density of states of \ce{Na_{$x$}TiO2} compounds in TB-I phase at different Na content $x$, calculated with the PBEsol XC functional.}
\end{figure}
Figure~\ref{fig7} displays the calculated PDOS of \ce{Na_{$x$}TiO2} compounds in TB-I phase at different Na content $x$.
It was found that the pristine \ce{TiO2} is a semiconductor with a wide bandgap of $\sim$3.0 eV, which is similar to the experimental value of rutile phase and smaller than that of the previous PBE calculation (3.88 eV)~\cite{Sicong}.
As in other transition metal oxides, its valence band maximum (VBM) is predominantly composed of O 2$p$ states, while the conduction band minimum (CBM) composed of Ti 3$d$ states.
Upon sodiation of \ce{TiO2} forming \ce{Na_{$x$}TiO2} compounds, such features of VBM and CBM were not severely altered, and moreover, Na states were not be found in these frontier bands, implying that Na atoms are fully ionized by donating their valence electrons to the host.
As increasing the Na content, the energy gap between the VBM and CBM decreases, and part of CBM became occupied states, demonstrating that the sodiation changes the compound from electron insulating to electron conducting material.
Since the original CBM is from the Ti 3$d$ states, it can be said that the electrons from the inserted Na atoms are transferred to the Ti atoms upon sodiation.

\begin{figure}[!th]
\centering
\includegraphics[clip=true,scale=0.09]{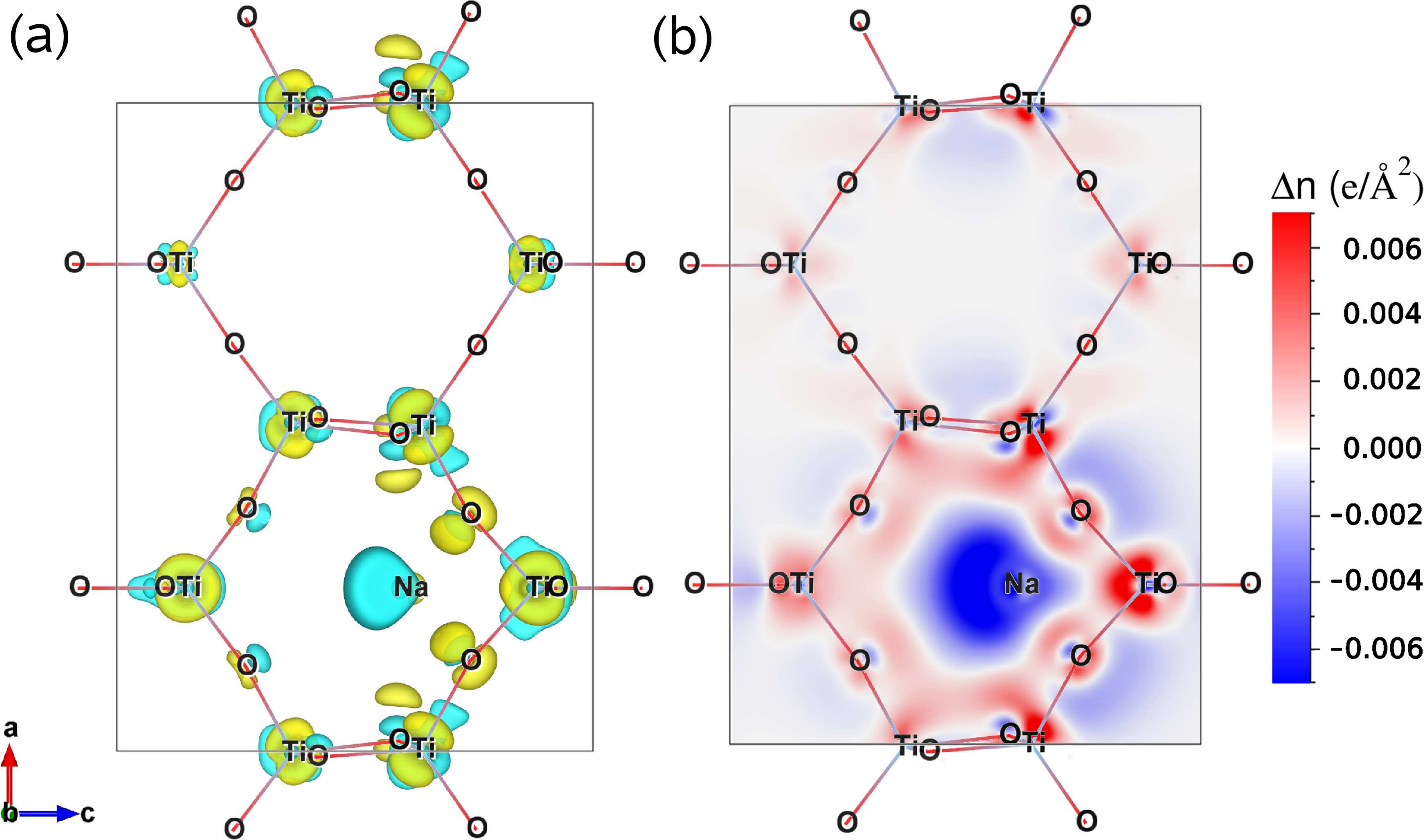}
\caption{\label{fig8}Electronic charge density difference in \ce{Na_{0.125}TiO2} compound in TB-I phase with (a) isosurface plot at the value of $\pm$0.0025 $|e|\cdot$\AA$^{-3}$, where yellow and blue colors denote the electron gain and loss, and (b) contour plot integrated over $b$-axis, where positive (red) and negative (blue) values indicate the electron gain and loss, respectively.}
\end{figure}
The electron transfer upon sodiation of \ce{TiO2} can be shown more intuitively by plotting the electronic charge density difference, defined as
\begin{equation}
\Delta n=n_{\ce{Na}_x\ce{TiO2}}-(n_{\ce{TiO2}}+xn_{\ce{Na}})
\end{equation}
where $n$ is the corresponding electronic density.
The electronic charge density difference  in \ce{Na_{0.125}TiO2} in TB-I phase is shown in Figure~\ref{fig8}.
It was observed that remarkable redistribution of electronic charge density occurs around mainly Na atom and nearby Ti and O atoms.
Through the electronic charge density difference integrated over $b$-axis, as shown in Figure~\ref{fig8}(b), it can be seen that there are charge depletion regions around the Na atom, and charge accumulation regions around the O atoms, while around the Ti atoms both the charge accumulation and depletion regions exist.
It can be thought that at the first step of \ce{TiO2} formation the Ti atoms give their 3$d$ electrons to the O atoms, becoming \ce{Ti^4+} and \ce{O^2-} ions, and at the second step of Na insertion the electron transfer occurs from the Na atom to the \ce{Ti^4+} ion, becoming \ce{Na+} and \ce{Ti^3+} ions.
This confirms that, as expected, Ti atoms play a redox couple of \ce{Ti^{3+}}/\ce{Ti^{4+}} upon insertion/extraction of Na atom.
Similar arguments can be found in TB-II, -III, and -IV phases (Figure S4).
Such remarkable electron transferring upon sodiation indicates the strong ionic bonding between ions and thus stable formation of \ce{Na_{$x$}TiO2} compounds in the TB phases.

\section{Conclusions}
In this work we provide an insightful understanding of electrochemical properties of sodiated \ce{TiO2} compounds in penta-oxygen-coordinated trigonal bipyramid phases, aiming at finding the potentiality of effective anode material of sodium-ion battery.
Identifying the four different crystalline structures as TB-I, -II, -III, and -IV phases, these structures were found to have comparable relative stabilities with the precedent \ce{TiO2} phases by comparing the DFT total energies calculated with the PBEsol XC functional.
Various supercells were constructed to make models of sodiated \ce{Na_{$x$}TiO2} compounds in these TB phases, including 8 \ce{TiO2} formula units and up to 8 Na atoms, with the Na content from $x=0$ to $x=1$ with an interval of 0.125, and their crystalline structures and energetics were investigated by using the PBEsol calculations.
The calculated sodium binding energies indicate the strong binding between Na and the host, and the exothermic reaction of Na insertion into \ce{TiO2} in TB phases.
Upon sodiation, TB-I and -IV phases were found to be expanded with the relative volume expansion rate of under 4\%, while TB-II and -III phases to be shrunk and expanded within the range of $\pm$10\%.
We also identified the intermediate structures lying on the convex hull lines by calculating the formation energies, and estimated the electrode voltage by using these structures, revealing that TB-I, -III, and -IV phases exhibit the low first step voltages around 1.0 V with specific capacity of $\sim$335 \scunit~but TB-II phase cannot be fully sodiated due to negative voltage from $x=0.5$.
Activation barriers for Na ion migrations were determined, indicating that the TB-I phase has the lowest activation barriers of under 0.35 eV at the Na content from $x=0.125$ to $x=0.5$.
Through the analysis of PDOS and charge density difference upon sodiation, moreover, TB-I phase changes from electron insulating to electron conducting material due to the electron transfer from Na atom to Ti ion. 
Based on these findings, TB-I among the four TB phases can be concluded to be the best choice for SIB anode material with reasonably low electrode voltage, high specific capacity and superior rate capability.
We want to emphasize that the high specific capacity of $\sim$335 \scunit~in particular is possible only by taking advantage of microporous and, at the same time, stable structural framework of unprecedented TB phases, which has never been achieved with other titania phases or any other materials based on rocking-chair mechanism.

\section*{Acknowledgments}
This work is supported as part of the fundamental research project ``Design of Innovative Functional Materials for Energy and Environmental Application'' (No. 2016-20) funded by the State Committee of Science and Technology, DPR Korea. Computations have been done on the HP Blade System C7000 (HP BL460c) that is owned by Faculty of Materials Science, Kim Il Sung University.

\section*{Appendix A. Supplementary data}
Polyhedral view for crystalline structures of pristine \ce{TiO2} compounds in TB-I, -II, -III, and -IV phases, isosurface plot of bond valance sum in these phases, table for structural properties of \ce{Na_{$x$}TiO2} compounds in TB-III and -IV phases, binding energy graphs of Na atom in these compounds as functions of Na content $x$, table for activation energies for Na ion migrations in these compounds at different Na contents, and contour plot of electronic charge density difference in these TB phases.

\section*{\label{note}Notes}
The authors declare no competing financial interest.

\bibliographystyle{elsarticle-num-names}
\bibliography{Reference}

\end{document}